%% file: ms.tex
\documentclass[manuscript, nonacm=true]{acmart}
\pdfoutput=1 
\AtBeginDocument{%
  \providecommand\BibTeX{{%
    \normalfont B\kern-0.5em{\scshape i\kern-0.25em b}\kern-0.8em\TeX}}}

\setcopyright{none}
\copyrightyear{}
\acmYear{}
\acmDOI{}

\acmConference[]{}{}
\acmBooktitle{ }
\acmPrice{ }
\acmISBN{ }

\usepackage{balance}       
\usepackage{graphics}      
\usepackage{color}
\usepackage{booktabs}
\usepackage{textcomp}
\usepackage{amsmath}
\usepackage{mathtools}
\usepackage{caption}
\usepackage{subcaption}
\usepackage{ccicons}          

\newif\ifShowNotes

\global\ShowNotesfalse

\ifShowNotes
\newcommand{\TODO}[1]{\textcolor{red}{\begingroup\raggedright TO DO: #1\\\endgroup}}
\newcommand{\NOTE}[2][gray]{\smallskip\noindent
  \colorbox{#1!30}{\parbox{.98\linewidth}{{\small\textbf{#2}}}}
}
\newcommand{\boldification}[1]{\textcolor{blue}{\begingroup\raggedright ** #1 **\\\endgroup}}
\else
\newcommand{\TODO}[1]{}
\newcommand{\NOTE}[2][gray]{}
\newcommand{\boldification}[1]{}
\fi

\newcommand{\MAYBECUT}[1]{}
\begin{document}



\title[Intersectionality Goes Analytical]
{Intersectionality Goes Analytical:\\
Taming Combinatorial Explosion Through Type Abstraction}

\author{Margaret Burnett}
\affiliation{%
  \institution{School of EECS, Oregon State University}
  \city{Corvallis}
  \state{Oregon}
  \country{USA}
  \postcode{97333}
}
\author{Martin Erwig}
\affiliation{%
  \institution{School of EECS, Oregon State University}
  \city{Corvallis}
  \state{Oregon}
  \country{USA}
  \postcode{97333}
}
\author{Abrar Fallatah}
\affiliation{%
  \institution{School of EECS, Oregon State University}
  \city{Corvallis}
  \state{Oregon}
  \country{USA}
  \postcode{97333}
}

\author{Christopher Bogart}
\affiliation{%
  \institution{Institute for Software Research, Carnegie Mellon University}
  \city{Pittsburgh}
  \state{Pennsylvania}
  \country{USA}
  \postcode{15213}
}

\author{Anita Sarma}
\affiliation{%
  \institution{School of EECS, Oregon State University}
  \city{Corvallis}
  \state{Oregon}
  \country{USA}
  \postcode{97333}
}

\renewcommand{\shortauthors}{Burnett et al.}

\begin{abstract} 

HCI researchers' and practitioners' awareness of intersectionality has been expanding, producing knowledge, recommendations, and prototypes for supporting intersectional populations.
However, doing intersectional HCI work is uniquely expensive: it leads to a combinatorial explosion of empirical work (expense 1), and little of the work on one intersectional population can be leveraged to serve another (expense 2).
In this paper, we explain how representations employed by certain analytical design methods correspond to type abstractions, and use that correspondence to identify a (de)compositional model in which a population's diverse identity properties can be joined and split.
We formally prove the model's correctness,
and show how it
enables HCI designers to harness existing analytical HCI methods for use on new intersectional populations of interest.
We illustrate through four design use-cases, how the model can reduce the amount of expense 1 and enable designers to leverage prior work to new intersectional populations, addressing expense 2.




\end{abstract}

\NOTE{Abstract max allowed wordcount: 150 words}




\maketitle


%

\newcommand{\set}[1]{\ensuremath{\{#1\}}}
\newcommand{\joinName}{\textit{Join}}
\newcommand{\join}[2]{\ensuremath{\joinName(#1,#2)}}


\newcommand{\dimFirstLetter}[1]{%
   \ensuremath{{\mathcal #1}\mkern-1.2mu}}
\newcommand{\dimFmt}[2]{%
   \ensuremath{\dimFirstLetter{#1}\textit{#2}}}
\renewcommand{\dim}{\dimFmt{D}{im}}
\newcommand{\Gender}{\dimFmt{G}{ender}}
\newcommand{\SES}{\dimFmt{S}{ES}}
\newcommand{\Age}{\dimFmt{A}{ge}}

\newcommand{\Facet}{\textit{Facet}}
\newcommand{\facet}{\textit{facet}}
\newcommand{\State}{\textit{State}}
\renewcommand{\state}{\textit{state}}
\newcommand{\Issue}{\textit{Issue}}
\newcommand{\spot}{\textit{spot}}
\newcommand{\persona}{\textit{persona}}

\newcommand{\Fmax}[1][{}]{\ensuremath{\max(\Facet_{#1})}}
\newcommand{\Fmin}[1][{}]{\ensuremath{\min(\Facet_{#1})}}
\newcommand{\spotMany}{\ensuremath{\overline{\spot}}}

\newcommand{\iMagName}{\textnormal{iMag}}
\newcommand{\iMag}[1][\dim]{\ensuremath{\iMagName(#1)}}
\newcommand{\xMag}{\textnormal{iMag}}

\input{content/1_introduction}
\input{content/2_background}
\input{content/3_related_work}
\input{content/4_operators}




\input{content/5_discussion}

\input{content/6_implications}
\input{content/7_conclusions}

%
%
%
%
%
\balance{}

\bibliographystyle{ACM-Reference-Format}
\bibliography{ms}

\end{document}

%% file: content/1_introduction.tex
\section{Introduction}

\begin{figure}
     \centering
     \begin{subfigure}[b]{0.33\textwidth}
         \centering
         \includegraphics[width=.5\textwidth]{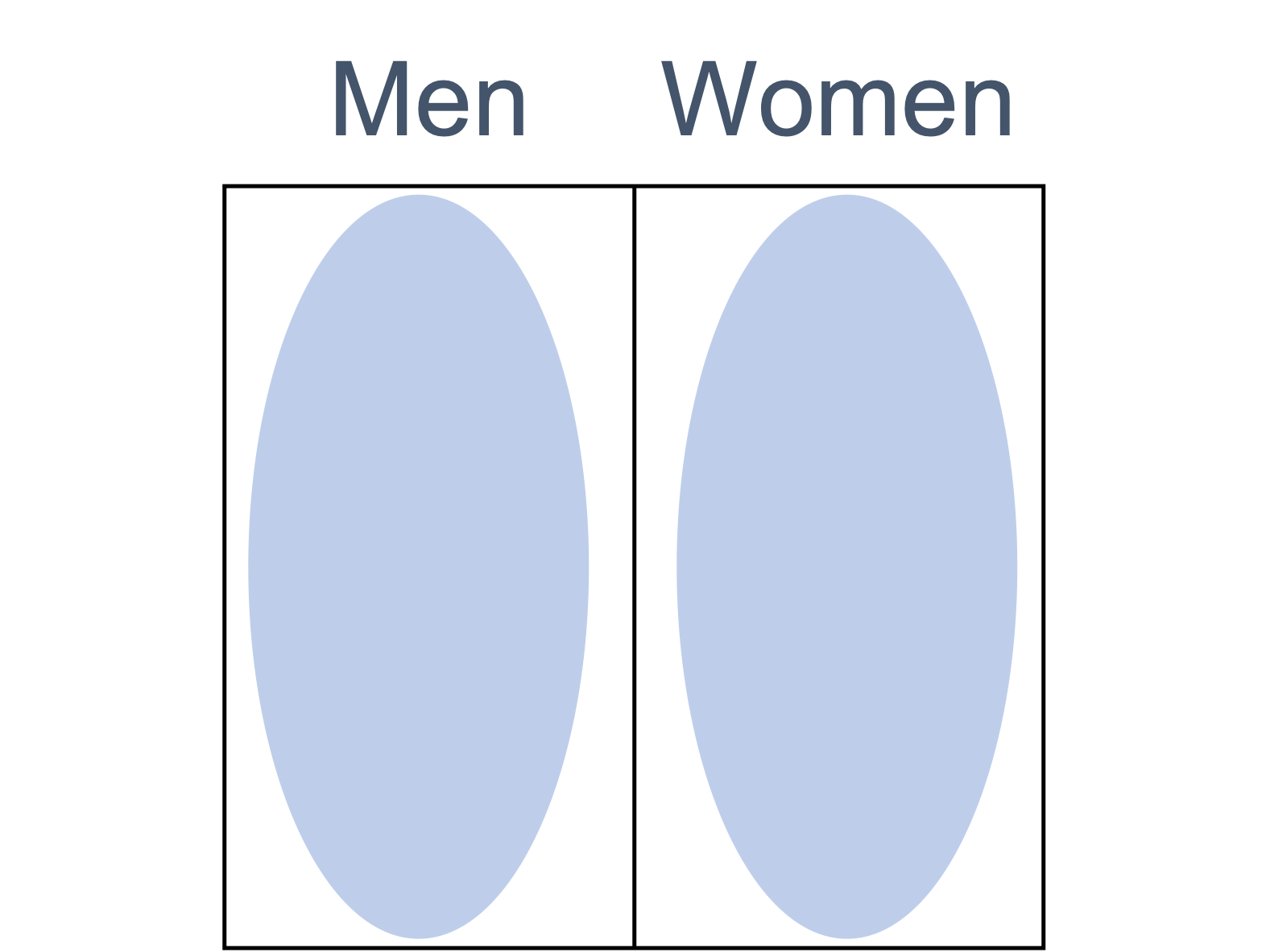}
         \caption{Gender}
         \label{fig:intersectionalityCells1}
     \end{subfigure}
     \begin{subfigure}[b]{0.33\textwidth}
         \centering
         \includegraphics[width=.5\textwidth]{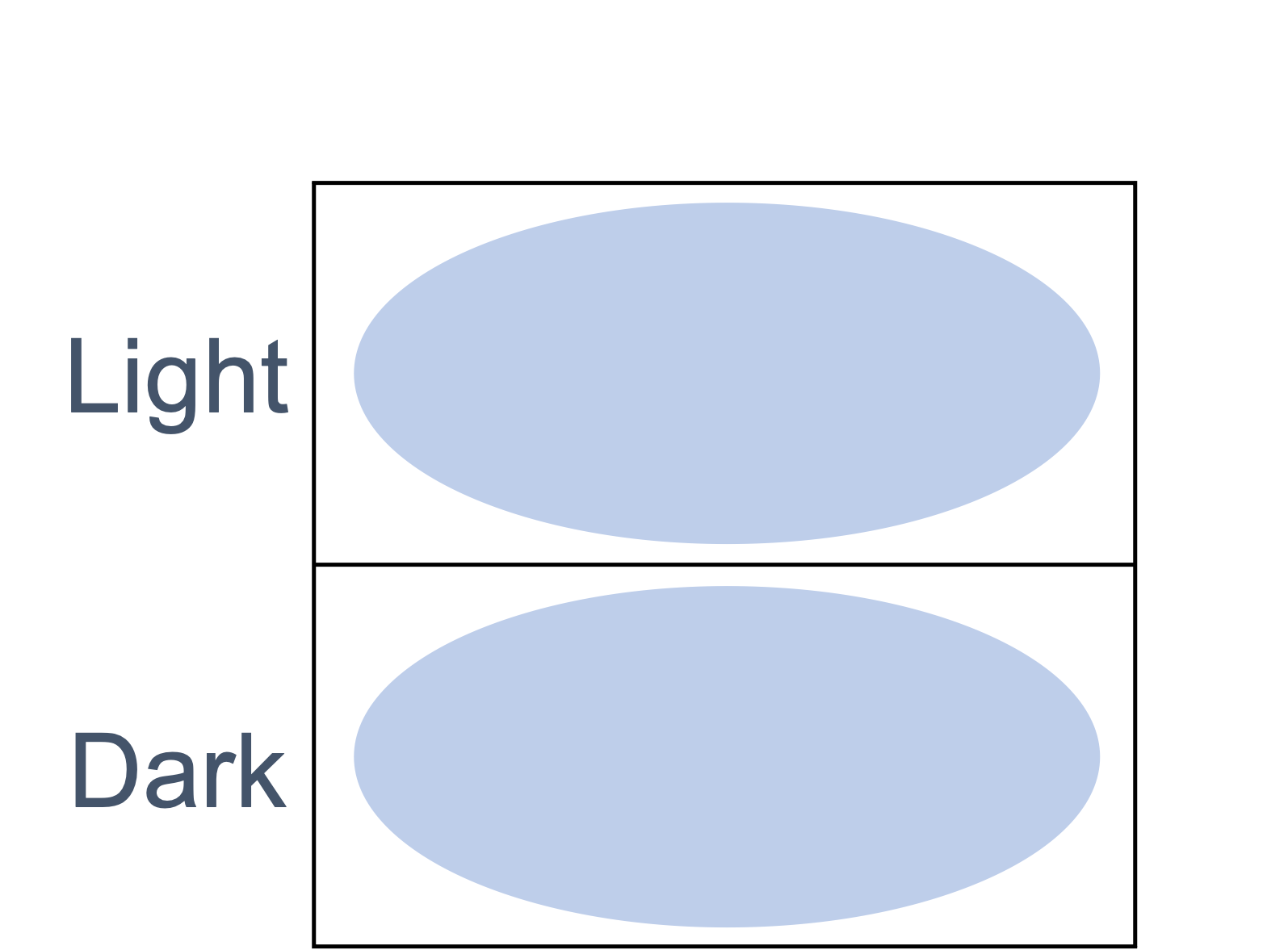}
         \caption{Skin color}
         \label{fig:intersectionalityCells2}
     \end{subfigure}
     \begin{subfigure}[b]{0.33\textwidth}
         \centering
         \includegraphics[width=.5\textwidth]{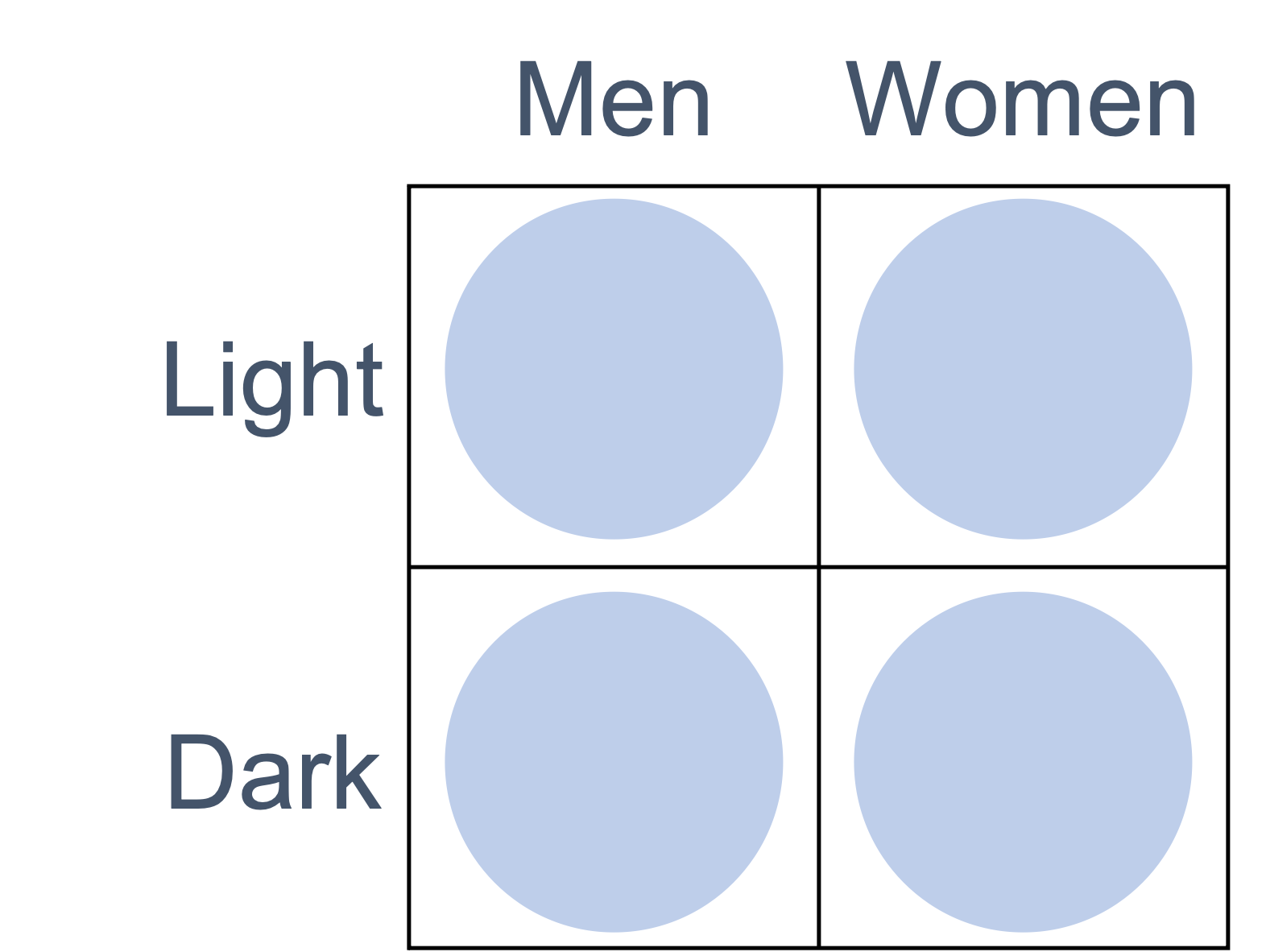}
         \caption{Gender and skin color}
         \label{fig:intersectionalityCells3}
     \end{subfigure}
        \caption{Example intersectionality issue over two diversity dimensions for facial recognition systems: gender and skin color.  Empirically analyzing a system's support of users by one dimension at a time (e.g., (a) first gender, then (b) skin color) can fail to reveal lack of support in some intersectional areas (e.g., (c) dark-skinned women)~\cite{buolamwini2017gender, buolamwini2018gender}.}
        \label{fig:intersectionalityCells}
        \vspace{-5mm}
\end{figure}

\boldification{Intersectionality is important.}
Practices in HCI have recently been shifting from designing for mythically homogeneous ``users'' towards more informed and/or disaggregated practices that recognize users' various and overlapping identities.
However, as a meta-review by Schlesinger et al.~\cite{schlesinger2017intersectional} revealed, much of this work considers only one identity dimension at a time, such as gender or race status.
Such one-dimensional approaches have been insufficient at understanding and supporting people with certain combinations of multiple, intersecting identities~\cite{crenshaw1989demarginalizing, rankin2020intersectional, rankin2020seat, erete2021can}.
A well-known example is the facial recognition failure rate for Black women~\cite{buolamwini2017gender,buolamwini2018gender}, in which facial recognition systems achieved high accuracy when predicting between gender (Figure~\ref{fig:intersectionalityCells1}), or skin color (Figure~\ref{fig:intersectionalityCells2}), but these one-dimensional comparisons did not reveal the huge lack of support for dark-skinned women (Figure ~\ref{fig:intersectionalityCells3}).



\boldification{, but designing for intersectional populations is UNIQUELY expensive}

However, addressing the intersectionality problem is uniquely expensive HCI work. 
At least two factors contribute to the expense.
First, intersectionality work has relied heavily upon  empirical work with real users in the particular population of interest (ranging from ethnography to statistical lab studies~\cite{costello2012increasing, booth2018diversifying, erete2021can}), which leads to the challenge of gaining access to and the willingness of  ``enough'' members of that particular understudied population to participate.
Even the single-dimensional approach of recruiting enough members across a single dimension of identity can be difficult, as evidenced by numerous empirical works with mostly men as participants despite significant efforts to recruit multiple genders. 

Second, most intersectional HCI work is not  scalable, because knowledge gleaned empirically about \textit{one underserved intersectional} population  does not easily leverage to \textit{other intersectional populations}.
Conducting a ``formative study---design---evaluate'' design process for each unique intersectional population produces a combinatorial explosion of unique HCI work needed.
As Schlesinger et al. summarize, ``Understanding users becomes increasingly complicated when we grapple with various overlapping attributes of an individual’s identity''~\cite{schlesinger2017intersectional}.


\boldification{the crux of what created this problem: underrepresentation in particular cells (Figure **, right). After all, nobody is worrying about problems facing light-skinned men in facial recognition software: because that intersectional population has contributed plenty of data and is well-represented in design decision-making. **D2.25**} 
At the heart of the need for intersectional HCI is the fact that some intersectional populations have much less representation than others in the decisions designers make. 
Returning to the facial recognition example, relatively few data points from the intersectional population of dark-skinned women were collected for the lower-right quadrant of Figure~\ref{fig:intersectionalityCells}(c)~\cite{buolamwini2017gender}, so that intersectional population had relatively little influence on the system's ultimate efficacy.
In contrast, the intersectional population of light-skinned men was overrepresented in the formative and in the evaluation data.
Not surprisingly, the system's efficacy for that intersectional population was very high~\cite{buolamwini2017gender}.
This suggests that a key to reducing the expense of serving intersectional populations is finding ways to give every intersectional population an equal voice---despite the challenges of getting access to an equal number of data points for each such population.

\boldification{TYPE ABSTRACTIONS to the rescue: TYPE abstractions can help. Here's what you need to know about types: (1)  types vs. values (programming and intersectionality) and (2) types are a synonym for "set of all possible values"... **D2**}
This is where \textit{type abstraction}, borrowed from programming language theory, can help.  
A \textit{type} in a programming language stands for  
\textit{all possible values} of that type. 
For example, type Integer stands for all possible integer values, and in the example of Figure~\ref{fig:intersectionalityCells}, type Skin Color stands for all possible skin color values.
Types thus provide equal representation to every value of a type.


This property of types provides an avenue for approaching intersectionality.  
Since an inclusivity method that uses \textit{type abstraction} to find gender-inclusivity issues gives equal representation to users of all genders, by the equal representation property of types; likewise, a type-based inclusivity method for skin colors gives equal representation to users of all skin colors.
Hence all four quadrants of Figure~\ref{fig:intersectionalityCells3} would now be equally populated. 
Thus, while the facial recognition algorithms described by Buolamwini~\cite{buolamwini2017gender} did not support dark-skinned women well due to unequal ``voice'' (data set size) for that population, a principled type-based approach would not have this weakness.

\boldification{BENEFIT 2: tractability.}
Beyond the benefit of equal representation, the second benefit of type abstraction is tractability, especially for certain ordinal types like self-efficacy or risk aversion used in HCI work (e.g.,~\cite{hekler2013theory, soden2020uncertainty}).  
%
Many such \textit{ordinal types}---types whose values have some kind of inherent order (e.g., risk-aversion values range from risk-tolerant to risk-averse; Skin Color values range from lighter to darker~\cite{fitzpatrick1988validity})---can be represented as a pair of two values: the ``smallest'' possible value and the ``largest'' possible value of that type. 
This brings tractability---human designers can now reason about a principally infinite set of values a type might have by looking at only a constant number of values.
%


\boldification{So, how CAN we reason in terms of types instead of instances -- Analytical methods **D2}
Harnessing the power of type abstraction in this way is an inherently analytical approach.\footnote{By \textit{analytical approaches}, we refer to inspection methods~\cite{nielsen1994inspectionMethods}, which HCI practitioners do \textit{without} directly using human data. Examples include cognitive walkthroughs, applying HCI guidelines to a design, and heuristic evaluations. Analytical methods are often used before empirical work with flesh-and-blood users, to catch and fix some design problems early in the design cycle, when fixing problems is less expensive than fixing them later in the cycle.}
Analytical methods, such as heuristic evaluation~\cite{nielsen1990heuristic} and cognitive walkthroughs~\cite{wharton1994, mahatody2010}, have contributed extensively to  HCI. 
They are complements, not replacements, for empirical work in HCI, because they can reduce the amount of empirical work needed to the sorts of things that \emph{only} empirical work can uncover.

\boldification{In this paper.. }


In this paper, we explain how an HCI designer can leverage the power of type abstraction and compose or decompose existing analytical methods and types in the InclusiveMag family of analytical methods~\cite{mendez2019inclusivemag}, to produce new analytical methods for new intersectional populations of interest.
We formally prove the (de-)compositionality properties of the analytical methods, and then illustrate their practical implications for HCI designers. 



\boldification{Our contributions are... }

The primary contribution of this paper is the first systematic approach for HCI practitioners to reason analytically about any intersectional population of interest, by systematically composing existing methods and/or type(s) that have been created for non-intersectional populations. The components of this contribution are:


\begin{itemize}
    \item An approach enabling HCI practitioners to reason about intersectional populations at the level of types instead of individual values.
    \item A (de)compositional model of identity traits in which diversity dimensions can be joined and split while preserving their analytical properties.
    \item An ``equal voice'' to every intersection of the identities the designer selects.
    \item A formal evaluation in the form of a formal proof of the model's correctness.
    \item Several practical usage scenarios for HCI practitioners in a variety of HCI design use-cases. 
\end{itemize}


\NOTE{MMB: Contribs list is the end of the Intro.  Please do not add anything below this line.}

%% file: content/2_background.tex
\section{Background}
\label{sec:bg}

\boldification{we're going to talk about InclusiveMag because...}
To reason at the level of type abstractions requires analytical methods consistent with the notion of types.
A type represents the set of all possible values of that type, so analytical methods that allow designers to reason about entire sets or ranges of diverse values fulfill this requirement.
One family of methods that fulfills it particularly well is the InclusiveMag family of analytical methods.

\boldification{inclusivemag is a meta-method to create inspection methods study specific diversity dimension.}
InclusiveMag~\cite{mendez2019inclusivemag} is a  meta-method that enables HCI researchers to generate systematic analytical design methods for a given diversity dimension. 
Designers and other software practitioners can then use the InclusiveMag-generated methods to evaluate user experiences from the perspective of users across the diversity dimension.

\begin{figure}[ht]
\includegraphics[scale=0.5]{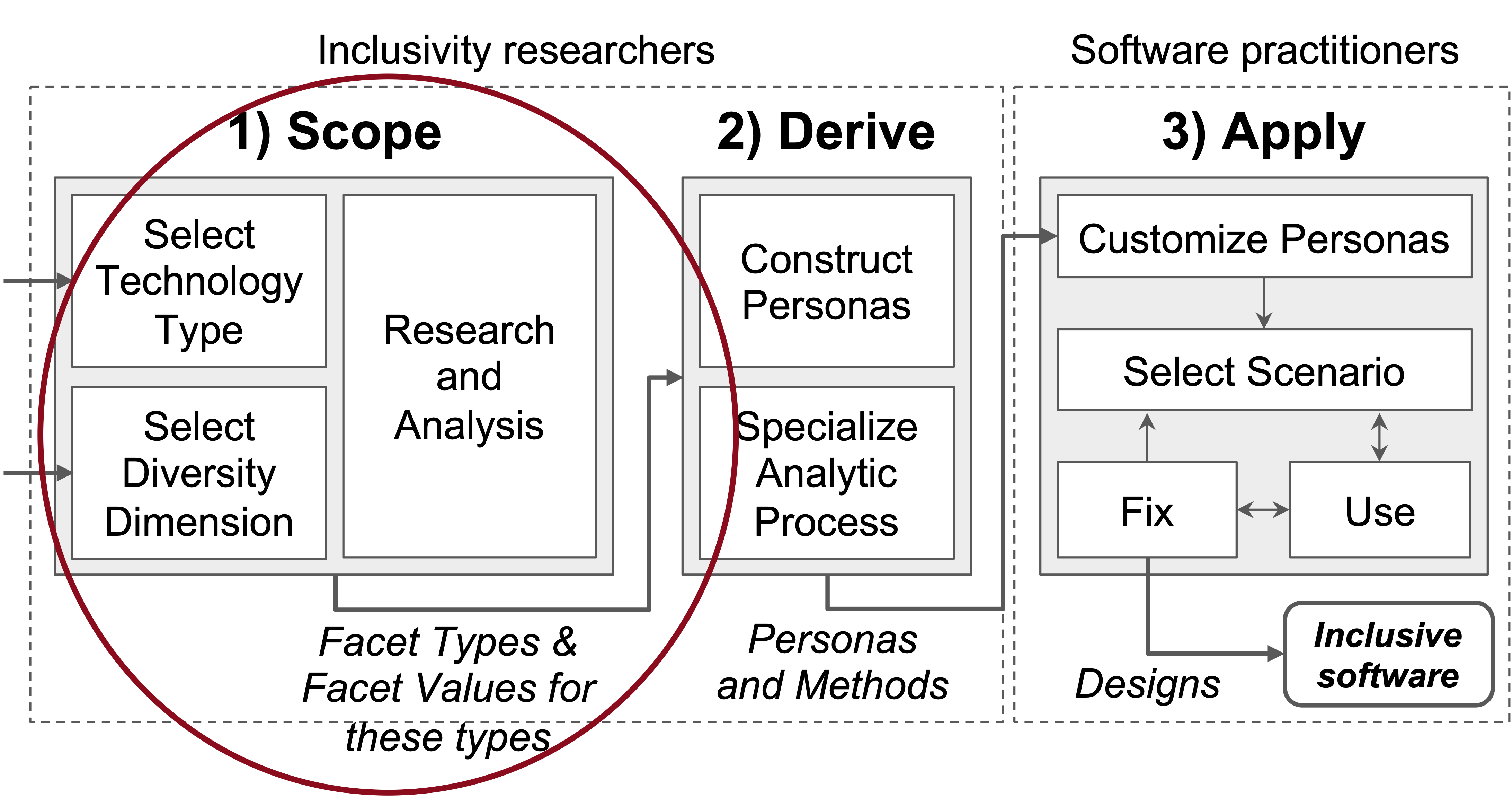}
\caption{The InclusiveMag meta-method has three steps. Step 1 (circled in red) is the one of interest to this paper.  Its output is facet types and the range of possible values that each facet type can have.  These facet types provide the starting point for the type-based reasoning we present in this paper.}
\label{figure:inclusiveMagProcessDiagram}
\end{figure}

\boldification{For example, it generated GenderMag and others}
For example, InclusiveMag was used to generate GenderMag, a systematic analytical method for the diversity dimension of gender~\cite{burnett2016gendermag-jrnl, mendez2019inclusivemag}.
HCI practitioners have used the GenderMag method  
to find and fix inclusivity issues in a variety of domains, such as education software~\cite{burnett2016gendermag-jrnl, Hilderbrand_2020gendermagPractices, shekhar2018gendermag-learningManagement, cunningham2016digitalLibrary}, machine learning aids
~\cite{burnett2016gendermagFieldstudy}, office productivity software~\cite{hill2017gendermagStereotyping}, Open Source project sites~\cite{chatterjee2021aid, ford2019beyond,padala2020genderOSS}, robotics~\cite{showkat2018gendermag}, software tools~\cite{gralha2019socialGoalModels}, and search interfaces~\cite{vorvoreanu2019gendermagEmpirical}.
Other offspring of InclusiveMag include SESMag to support users in diverse socioeconomic situations~\cite{Hu2021Toward}, AgeMag to evaluate age bias in e-commerce applications~\cite{mcintosh2021agemag}, and a collection of eight pilot InclusiveMag-generated methods to support eight diversity dimensions (e.g., eyesight,  attention span, position along the autism spectrum)~\cite{mendez2019inclusivemag}.

\NOTE{ME: Replace ``<diversityDimension>Mags''.}

\begin{figure} [hb]
\includegraphics[scale=0.8]{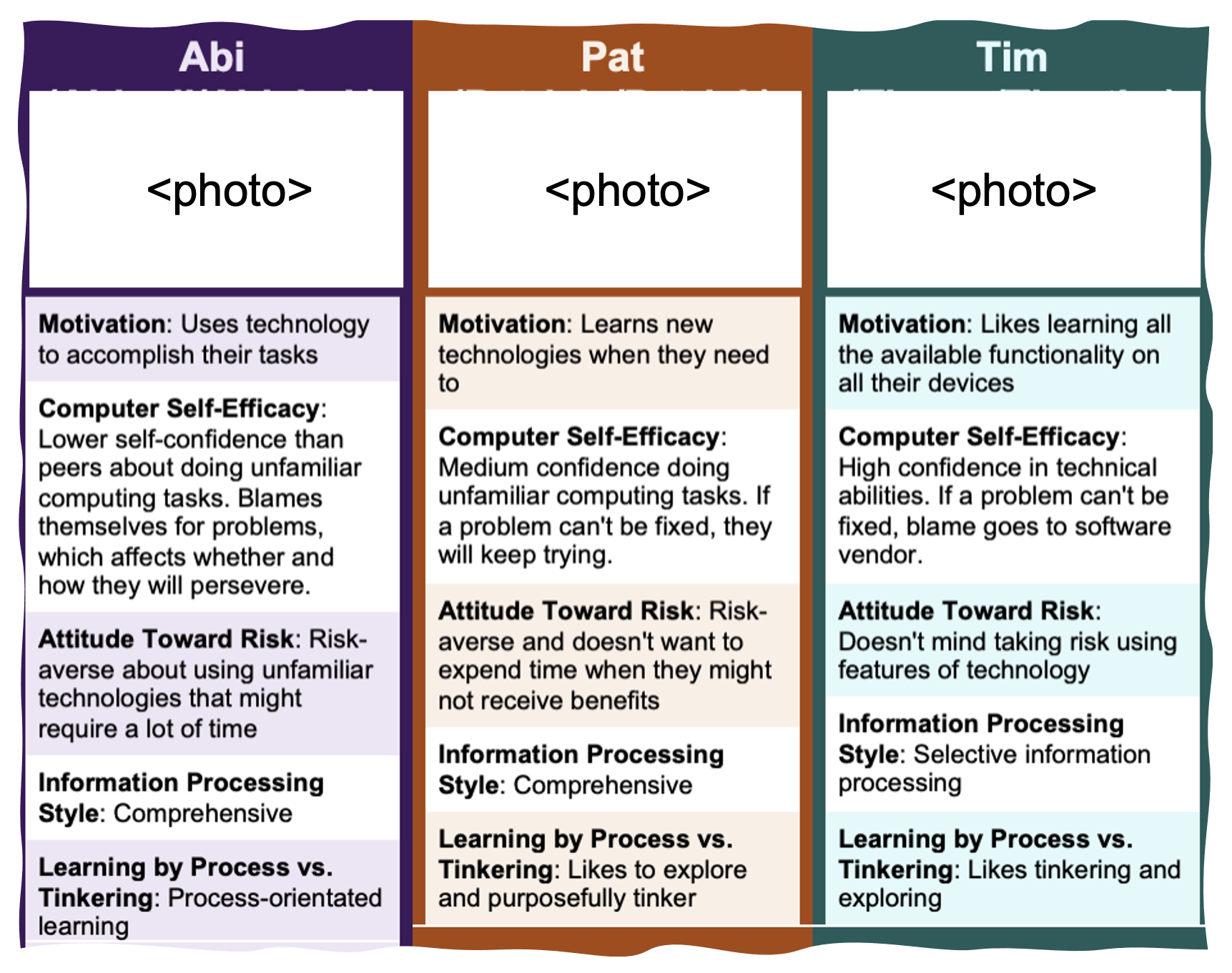}
\caption{Portions of GenderMag's three personas---Abi (left), Pat (middle), and Tim(right)---as customized by a faculty member who was applying GenderMag to college-level students~\cite{letaw2021online}, with each persona's facet value for the GenderMag facet types. (From a type-based perspective, two personas are sufficient to capture the two endpoints of each facet type's range, but  more than two personas is useful in emphasizing to other humans the diversity of the target population.)}
\label{fig:GMpersonas}
\end{figure}

\boldification{InclusiveMag scope step is the one that produces the types we care about.}
InclusiveMag's Scope step, circled in Figure~\ref{figure:inclusiveMagProcessDiagram}, produces a set of facet types for the diversity dimension of interest (e.g., gender, in the GenderMag example).  
These \textit{facet types} represent traits for which individuals at opposite ends of the diversity dimension can differ significantly from each other.
For example, the GenderMag personas in Figure~\ref{fig:GMpersonas} enumerate GenderMag's five facet types and some of the possible facet values different individuals might have.  
This paper uses such facet types for the type-based reasoning presented here.

\NOTE{-----------MMB: THIS IS A FIREWALL.  BELOW THIS LINE, always say FACET TYPES/VALUES, not traits, not attributes, not factors, etc.-----------}

\boldification{InclusiveMag's second step takes those facet types and uses them to produce useful stuff for designers. And in the third step, designers actually use them.}
%
%
In InclusiveMag's second step, inclusivity researchers use the facet types they created in the first step to derive mechanisms for HCI practitioners to use when designing/evaluating a system's inclusivity, such as Figure~\ref{fig:GMpersonas}'s personas.
The researchers also specialize an existing analytic method, such as a cognitive walkthrough or set of design heuristics, using the facet types. 
Finally, in the third step, HCI/software practitioners customize and actually apply the generated method(s) or other facet-based artifacts (e.g., personas) to evaluate/redesign their technology to increase its inclusivity across that diversity dimension.
This paper shows how these practitioners can compose one or more of these methods and/or sets of facet types to be able to evaluate/redesign their technology to be more inclusive across multiple, intersecting, diversity dimensions.

\NOTE{MMB: above sentence "This paper shows" should be the last line of this section.  Please do not add anything below this line.}

%% file: content/3_related_work.tex
\section{Related Work}
\boldification{What is Intersectionality and why it matters}
Kimberlé Crenshaw introduced and defined \emph{intersectionality} as a theoretical approach to illustrate how the interconnected nature of social categorizations such as gender, race and class creates an overlapping and independent system of discrimination or disadvantages ~\cite{oed:intersectionality, crenshaw1989demarginalizing}.
Crenshaw defined intersectionality to explain the limitation of single-axis analysis on antidiscrimination legislation and antiracist policies ~\cite{crenshaw1989demarginalizing}. 
Crenshaw's example to explain the term is the DeGraffenreid v. General Motors lawsuit, where five Black women alleged that General Motors' ``last-hired, first-fired'' lay-off policy perpetuated discrimination practices ~\cite{1976degraffenreid, crenshaw1989demarginalizing}. 
However, the Court's research failed to identify such discrimination, because General Motors had hired white women and Black men and the research neglected the complained race and gender discrimination that Black women experience ~\cite{1976degraffenreid, crenshaw1989demarginalizing}.  
Since then, the term intersectionality has been used in various fields, including Information Technology and Human-Computer Interaction (HCI)~\cite{ames2011understanding, buolamwini2018gender, ross2020intersection, rankin2020intersectional}. 

\boldification{Insersectionality definition in HCI}
In HCI, \emph{intersectionality} is a framework that considers how complex identities, situations, backgrounds, and experiences shape the contextual surroundings of underrepresented individuals~\cite{erete2018intersectional, schlesinger2017intersectional, rankin2020intersectional, Villacres2018Designing}.
Intersectionality implies that many underserved individuals have unique experiences, and that these experiences cannot be unraveled by considering a single social categorization or diversity dimension~\cite{erete2018intersectional, buolamwini2017gender, rankin2020intersectional}.
For example, the disparities in the accuracy of facial recognition algorithms classifying darker-skinned females cannot be unraveled by considering either gender or skin color alone, but rather the combination of gender and skin color~\cite{buolamwini2018gender}.

\boldification{So here's where the intersectional research in HCI lies...}

Research to understand intersectionality in software design falls into three categories. 
(1) Empirically studying the need for intersectional HCI, e.g., via ethnographic or qualitative studies with members of the population of interest~\cite{costello2012increasing, booth2018diversifying, mcfarlane2020get, rankin2020seat, erete2021can, Prana2020Including}. This is by far the most frequent in the literature.
(2) Developing and evaluating concrete software products that aim to address the needs of some intersectional population data (e.g., race and gender AI training data set)~\cite{Russis2020Data, buolamwini2018gender, buolamwini2017gender}. This has been sparsely studied in literature.
(3) Examining the complexity of the experiences across a class of software and any arbitrary intersectional population of interest~\cite{smyth2014anti, rankin2020intersectionality}. This is by far the least studied and is the category in which our work falls.


\boldification{category 1}
\textit{Category 1: Empirical studies of the need for intersectional research.}
Studies in this category intentionally recruit individuals from marginalized social categories or individuals with intersectional identities to shed light on their lived experiences.
Research into the efficacy of computing outreach activities for minority or intersectional students (high school or undergraduate) is particularly common. 
For example, ``Get Paid to Program'' is an after-school program that teaches low-income, high-school women to code, and has been shown to increase students' computing self-efficacy and refine students' career interests~\cite{mcfarlane2020get}.
Other examples of such outreach activities that have been investigated include Grace Hopper Scholars and the Glitch Game Tester~\cite{costello2012increasing, disalvo2013workifying}. 

\begin{figure}[hb]
    \centering
      \includegraphics[width=0.7\textwidth]{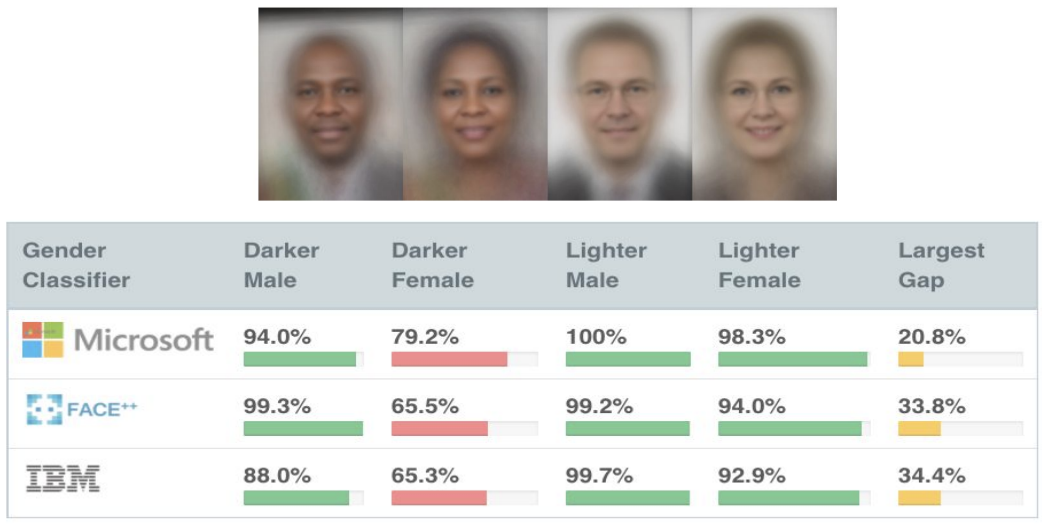}
  \caption{Reporting the accuracy of facial recognition classifiers (i.e., Microsoft, Face ++, and IBM) along social categories (i.e., skin color and gender) is significant to highlight the disparities within subgroups. Note that darker-skinned individuals are less likely to be classified than lighter-skinned individuals, and darker-skinned females are less likely to be classified than lighter-skinned males and females.~\cite{Buolamwini2019WrittenTestimony}}
    \label{fig:Buolamwini_Joy}
    \vspace{-3mm}
\end{figure}
Another approach in this category is a form of experience sampling~\cite{larson2014experience, csikszentmihalyi2014validity}, i.e., collecting experiences of individuals with intersectional identities~\cite{TuliS2019Its, Ismail2019empowerment}.  
A recent example of this type is the 2021 paper by  Erete et al.'s experience sampling paper~\cite{erete2021can}. 
Set in the current climate of the COVID-19 pandemic and escalating attention to systemic racism, the authors' experiences help raise awareness in the HCI and CSCW community of the systemic oppression Black feminist scholars endure~\cite{erete2021can}.
The authors' unique experiences are deeply rooted in their intersectional identities: (1)~being a Black feminist in an environment where black and brown communities are disproportionately affected by a worldwide pandemic, COVID-19, 
(2)~enduring years of systemic racism and police brutality in the U.S.,
and (3)~being a Black woman in a dominantly white research field that historically marginalized and erased their experiences~\cite{erete2021can}.

A third approach is the ``single-axis'' work; i.e., work that was originally carried out in a non-intersectional way that revealed the need to carry out such work in an intersectional way instead. For example, to examine the impact of single-axis interventions aiming to serve non-intersected but underserved groups, Rankin et al.\ interviewed 14 Black women at various levels in their CS careers~\cite{rankin2020intersectional}.
The work concluded that gendered-focused interventions had negatively impacted recruitment and retention of Black women, especially in predominantly white institutions that do not foster mentoring or supportive learning~\cite{rankin2020intersectional}.

\boldification{category 2}
\textit{Category 2: Concrete software products.}
Research in this category focuses on developing and/or evaluating a concrete software product for an intersectional population.  
A notable example of this category is Joy Buolamwini's Ph.D. dissertation featuring the creation and evaluation of Pilot Parliaments Benchmark (PPB), an inclusive benchmark training dataset for intersectional facial recognition algorithms~\cite{ buolamwini2018gender, buolamwini2017gender}. 
To create PPB, Buolamwini gathered 1270 images to populate each gender/skin color subgroup equally.
In conjunction with the usual binary gender classification (Female vs. Male), Buolamwini labeled the PPB using the Fitzpatrick skin color classification system~\cite{fitzpatrick1988validity}, which has six categories: lighter I-III and darker IV-VI, depending on melanin pigmentation.
She then evaluated the accuracy of three AI classifiers (Microsoft, Face++, and IBM) intersectionally using the PPB dataset. She showed that giving attention to equal representation in training data in each intersectional group resulted in higher accuracy than what other datasets had been able to achieve~\cite{buolamwini2017gender, Buolamwini2019WrittenTestimony}. 
She emphasizes, that at a minimum, we need to report the accuracy of intersectional subgroups because ``we cannot assume data collected from one demographic group can be extrapolated to other groups. Even within a demographic group, we need to account for intragroup variation''~\cite{buolamwini2017gender, Buolamwini2019WrittenTestimony}. 
Buolamwini was one of the first to reveal intersectional differences in accuracy of AI facial recognition algorithms, which tended to be much lower for dark-skinned women than for either women or dark-skinned people (Figure~\ref{fig:Buolamwini_Joy}).

\boldification{category 3}
\textit{Category 3: Anlaytical studies.}
The last category of intersectional HCI research analytically studies the oppressions, misconceptions, and fallacies that are upheld when designing software for users with intersectional social identities. 
This category relies on oppression and intersectionality theories by Patricia Hill Collins and Kimberle Crenshaw, and is profoundly rooted in social work and women's studies. 
An example of this category is the Anti-Oppressive Design Framework introduced by Symth et al.\ to guide the understanding and responses to the complexity of the experiences of intersectional users~\cite{smyth2014anti} in both the things we create and the environments in which we design. 
Current intersectional-HCI analytical studies center around several questions that individuals at the decision-making table should ask, such as: What oppression would this work strive to eliminate? At what level? and At which intersections? ~\cite{smyth2014anti}
Am I thinking about everyone as I am designing this tool? Who is at the table to inform decisions about how this technology will be designed? and for whom?~\cite{rankin2020intersectionality}. 
Rankin et al.\ stress the importance of these probes, especially within the HCI community, and when striving for inclusivity~\cite{rankin2020intersectionality}.

\boldification{Sunset: this last one one is the closest to our approach.}


Approaches like Smyth et al.'s and Rankin et al.'s~\cite{smyth2014anti, rankin2020intersectionality} are still rare.
Unlike approaches that investigate problems and needs tied to lack of support for intersectionality, these two works emphasize potential intellectual tools for designers to avoid at least some such problems.
Our work falls into this category, with a step-by-step process for designers to leverage existing analytical methods to support intersectionality.

%% file: content/4_operators.tex
\section{Type-Based Analytical Methods for Intersectionality}
\label{sec:formal}

\NOTE{ME: I think we shouldn't use the term \emph{intersectionality operator(s)}, since the operators are more general/generic. They just happen to support intersectionality. I've renamed the section.}


\NOTE{MMB to everyone -- CHI readers like to read about PEOPLE.  Two points about this: (1) Insert people-words rather than abstract phrases when possible (eg, instead of "members" possibly say "people"); (2) PEOPLE <> PERSONAS.  It's great to write about people and it's not the same thing as writing about personas.}

\boldification{Our aim is to contribute an ANALYTIC method that is SYSTEMATIC.  The InclusiveMag family provides an entrypoint for us to do so.}

Our aim is to enable HCI designers to create, just in time, their own analyses to systematically consider ``inclusivity bugs'' for  users with as many intersecting identities as the designer chooses to consider.  
The InclusiveMag family of methods provides an entry point for us to do so.
It uses extreme values of each population trait (e.g., high and low self-efficacy) to give HCI designers a process by which to ensure that the software simultaneously supports individuals at each end of each trait's range of values.  
These traits are chosen with the assumption that it is sufficient to design for the two endpoints of the scale, such as for people with both high and low technology self-efficacy,
since solutions that \emph{simultaneously} support both ends of the ordered range also support values within the range.\footnote{Some exceptions to this assumption exist. For example, testing f(x) = 1/x for MAXFLOAT and MINFLOAT misses  the problem at 0. When facet types have ``special values'' that violate this assumption, the method proposed here will miss such special values.}

\boldification{How?}
How can we apply InclusiveMag to a multidimensional population of possible identities?  
%
Prior intersectionality works have shown the insufficiency of considering, for example,
gender and socioeconomic status separately. 
But to specifically consider
a sufficient sample of all the possible facet value combinations a person could have, one would have to consider the needs of on the order of $2^N$  kinds of people to handle $N$ traits.
Figure~\ref{fig:AnalyticalPower} illustrates the problem; if a design team who can afford to do a fixed number of investigations tries to consider the needs of individuals with many facet values across many dimensions, their sampling becomes less and less adequate, spread more and more thinly through the multi-dimensional space of possible traits. 



\boldification{LINEAR: People have identities, but software has features that impact facet values, that can be
analyzed independently}
InclusiveMag-generated methods address the expense of the need for multiple investigations with people with different traits (Figure~\ref{fig:AnalyticalPowerA}) by: (1) identifying facet types (i.e., range of possible values of a human trait) that tend to differ (statistically) along some identity spectrum, and (2) eliminating inclusivity bugs that affect people at both ends of each facet type's range of possible values.  The intersectional approach we describe below extends this technique to multiple dimensions (Figure~\ref{fig:AnalyticalPower}b,c). 
In the following section, we describe the space of intersectional design in terms of sets of facet types, and show that adding dimensions of intersectionality means adding, not multiplying, facet types to consider, making intersectional analysis tractable for a design team with limited resources.  


\newcommand{\subfigsize}{0.33}
\newcommand{\imgfrac}{0.49}

\begin{figure}
     \centering
     \begin{subfigure}[b]{\subfigsize\textwidth}
         \centering
         \includegraphics[width=\textwidth]{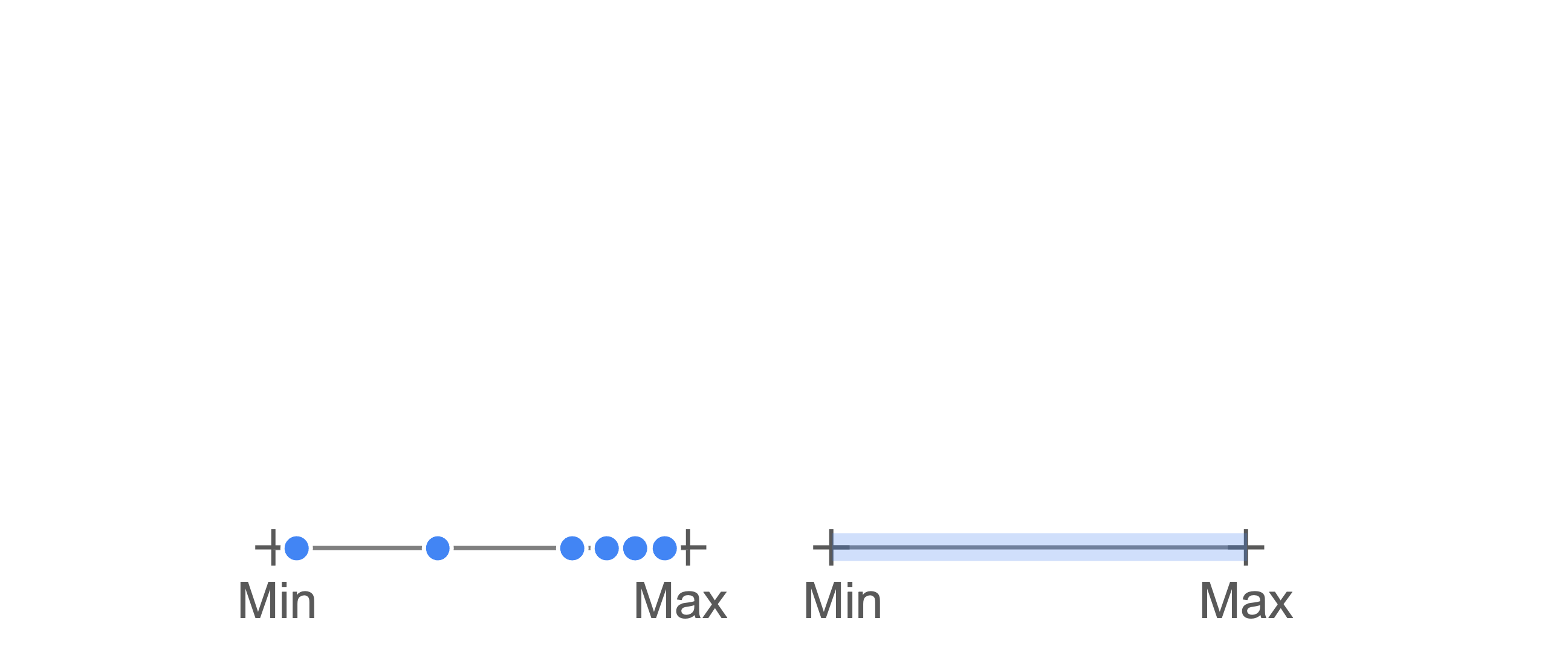}
         \caption{Single Dimension}
        \label{fig:AnalyticalPowerA}
     \end{subfigure}
     \begin{subfigure}[b]{\subfigsize\textwidth}
         \centering
         \includegraphics[width=\textwidth]{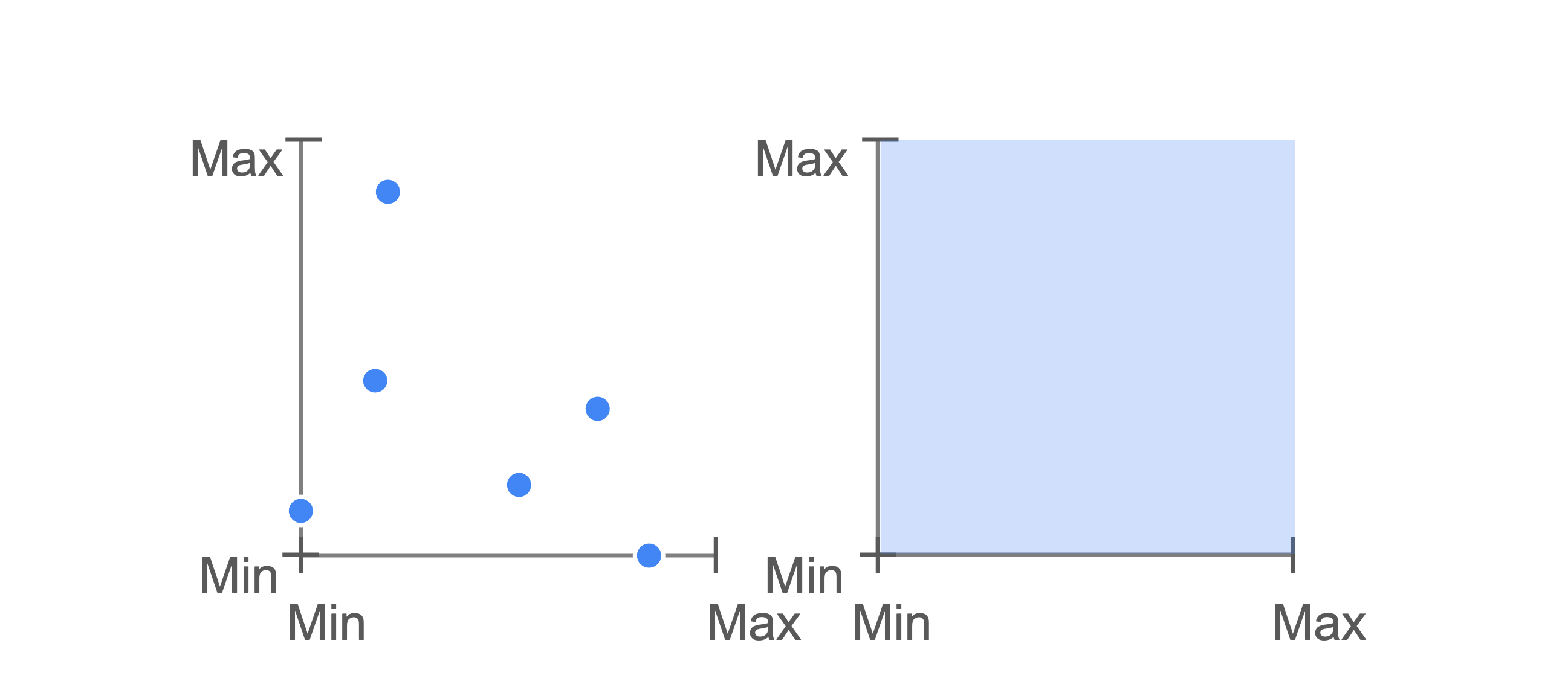}
         \caption{Two Dimensions}
         \label{fig:AnalyticalPowerB}
     \end{subfigure}
     \begin{subfigure}[b]{\subfigsize\textwidth}
         \centering
         \includegraphics[width=\textwidth]{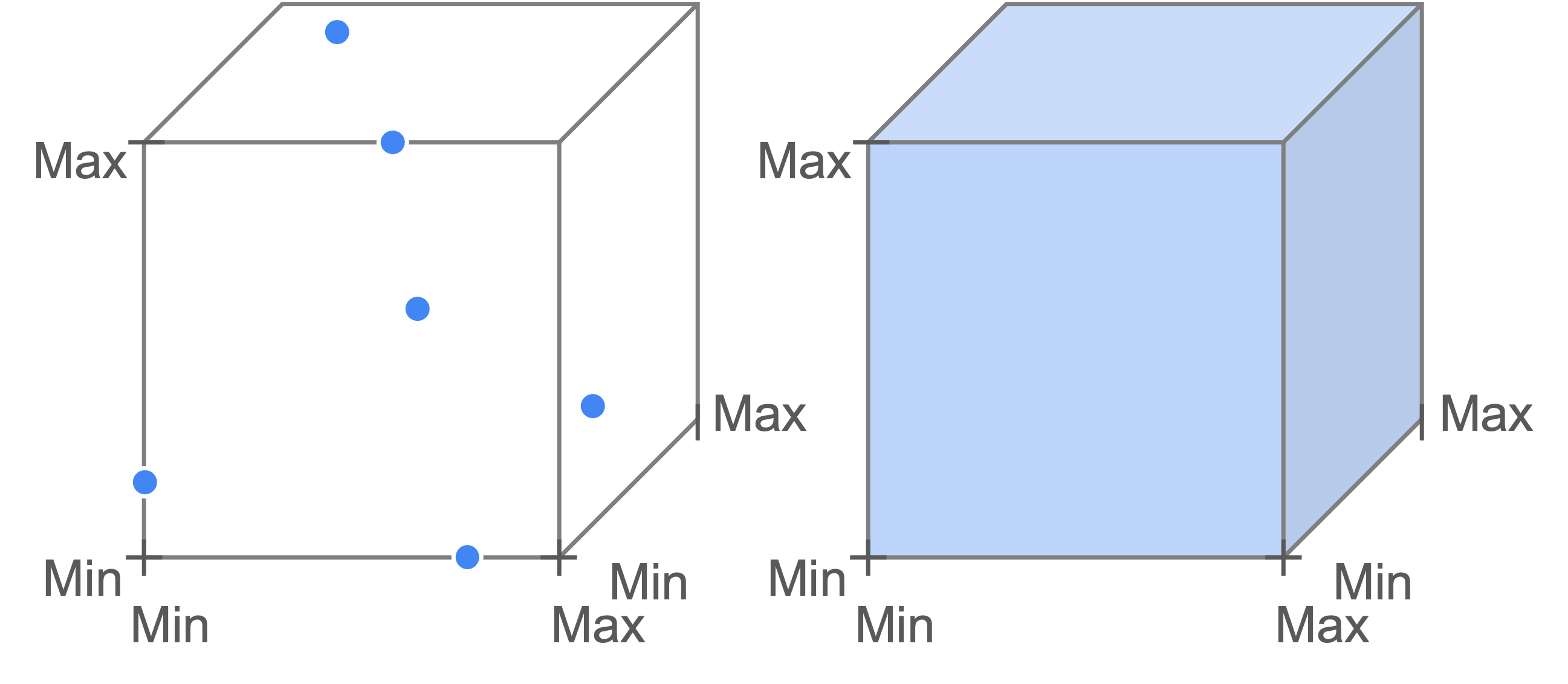}
         \caption{Three Dimensions}
         \label{fig:AnalyticalPowerC}
     \end{subfigure}
        \caption{The analytical power of a type-based approach.  A design team's effort is spread more and more thinly as they spend limited resources to explore an exponentially-growing design space. Types provide equal coverage along each dimension and full coverage of the space.   
        }
        \label{fig:AnalyticalPower}
\end{figure}





\subsection{Formalization}

We begin by formalizing the notions of \emph{diversity dimensions} along which people may fall (e.g., gender, socioeconomic status, etc.), and \emph{facet types}, which define fine-grained behavioral spectra (e.g., a spectrum of people's attitudes toward technological risk, a spectrum of their preferred information processing styles, etc.). 
We use the notational convention that \emph{types}---which are equivalent to \emph{sets} of values---start with a Capital letter, and \emph{values} (instances of those types, elements of those sets) start with a lowercase letter. \emph{Dimensions}, which are sets of facet types, start with a \dimFirstLetter{C}alligraphy capital letter.


As explained in Section \ref{sec:bg}, the InclusiveMag meta-method, which we will abbreviate as \iMagName, generates for a given diversity dimension \dim\ an analytical inclusivity method \iMag. 
Thus, \iMagName\ can be viewed as a function parameterized by a diversity dimension that yields an inclusivity method for a specific dimension.
For example, GenderMag is generated by applying \iMagName\ to $\dim = \Gender$, so $\textrm{GenderMag} = \iMag[\Gender]$.




The first step in \iMagName\ reduces the diversity dimension to a set of facet types (see ``Scope'' in Figure~\ref{figure:inclusiveMagProcessDiagram}). We formalize this by defining each \dim\ to be a set of facet types, that is, $\dim=\set{\Facet_1,\ldots,\Facet_n}$. Each facet type \Facet\ is an ordered set of all possible values of that type, whose minimum and maximum values are denoted by \Fmax\ and \Fmin, respectively.


The purpose of \iMag\ is to analyze a software use case's inclusivity across the diversity dimension \dim. 
\iMag\ works by examining a set of states $\State=\set{\state_1,\ldots,\state_n}$ of the software use case for two contrasting, extreme values, for each of the facet types.
These contrasting extreme values are usually captured by personas.
To cover the complete space of all facet values, the two personas are designed so that they each represent the opposite extreme values for every facet type.
A \dim-persona is thus defined as a set of facet values \set{\facet_1,\ldots,\facet_n} with $\facet_i\in\set{\Fmax[i],\Fmin[i]}$.

The designers' task is now to use their \iMagName-generated method to find inclusivity issues in their software. They examine a state $\state\in \State$ through the lens of an extreme facet value--\Fmin[i] or \Fmax[i]--and then potentially identify one or more issues of type \Issue.
This single analysis step can be formally represented by a function $\spot:\Facet\times\State\to 2^{\Issue}$ that takes a facet value and a state and returns a set of issues.\footnote{The notation $f:A\to B$ says that function $f$ has the type $A\to B$, that is, $f$ takes arguments of type $A$ and returns values of type $B$. $\Facet\times\State$ denotes the set of all pairs of a facet value and a state, and $2^{\Issue}$ denotes the powerset of \Issue.  (Note that \spot\ returns a set of issues and not individual ones. So if there are $N$ possible issues, then there are $2^N$ possible sets of issues.)}
For a given persona $\persona=\set{\facet_1,\ldots,\facet_n}$, \iMag\ applies $\spot(\facet,\state)$ for each facet value $\facet\in \persona$ and state $\state\in \State$.\footnote{There is no technical need to use personas: Instead of iterating \spot\ over two personas $\persona_1$ and $\persona_2$, one can just as well apply \spot\ to both extreme facet values, \Fmax\ and \Fmin, in each \Facet\ of the personas.}




%
To simplify the following definition, we define the function \spotMany, which identifies all issues for both extreme facet values of a facet type and a particular state.
\[
\spotMany(\Facet,\state) = \spot(\Fmin,\state)\cup\spot(\Fmax,\state)
\]
With \spotMany\ we can now formally define the function \iMagName, which collects for a dimension \dim\ and a set of states \State\ all issues identified by \spotMany. It does so by applying \spotMany\ to all combinations of state and extreme facet values and taking the union of all the results. Note that, when a method is actually carried out for some use case, we parameterize \iMagName\ not only by \dim, but also by \State, the set of states in that use case.
\begin{equation*}
\iMag[\dim,\State] =
\bigcup_{\mathclap{\substack{%
\Facet\in\dim \\
\state\in\State}}}
\spotMany(\Facet,\state)
\label{def:iMag}
\end{equation*}
%
%
To apply the meta-method \iMagName\ to the union of two dimensions means to apply \iMagName\ to the facet types from both dimensions, i.e., the union of both dimensions' facet type sets.
We call this union operation on dimensions, the \emph{join} of two dimensions, since this way of combining dimensions provides a joint view of two diversity dimensions.
\begin{equation*}
\join{\dim}{\dim'} = \dim\cup\dim'
\label{def:join}
\end{equation*}
The join of two dimensions represents the entire space of the two dimensions, including their intersection. 



\subsection{Compositionality Theorem and Proof}

To analyze an intersectionality population characterized by two diversity dimensions \dim\ and $\dim'$, we could apply \iMagName\ directly to the joint dimension \join{\dim}{\dim'}.
But we can just as well combine, that is, take the union of, the result of two independent runs of \iMagName\ for the two dimensions \dim\ and $\dim'$, since the following relationship holds.

\newtheorem{thm}{Theorem}[section]
\begin{thm}{(Compositionality of \iMagName)}
\[
\iMag[\join{\dim}{\dim'},\State] = 
\iMag[\dim,\State]\cup \iMag[\dim',\State]
\]
\label{th:comp}
\end{thm}

\noindent
In the proof of the theorem we make use of the following property of set union that follows directly from the associativity of set union: 
%
\begin{equation*}
\bigcup_{\mathclap{x\in A\cup B}} f(x) =
\bigcup_{\mathclap{x\in A}} f(x) \cup 
\bigcup_{\mathclap{x\in B}} f(x)
\label{lem:assoc}
\end{equation*}
Now we can prove the theorem as follows.

\newlength{\saveskip}
\setlength{\saveskip}{\abovedisplayskip}
\setlength{\abovedisplayskip}{-2.7ex}

\medskip\noindent
\textbf{Proof.}
\begin{align*}
&\iMag[\join{\dim}{\dim'},\State] \\
&= \iMag[\dim\cup\dim',\State]     
& (\textrm{Definition of \emph{Join}})  \\
&= \bigcup_{\substack{\Facet\in\dim\cup\dim' \\ \state\in\State}}
   \spotMany(\Facet,\state)
& (\textrm{Definition of \iMagName}) 
\\[1ex]
&= \bigcup_{\substack{\Facet\in\dim \\ \state\in\State}}
\spotMany(\Facet,\state)
\quad\cup
\bigcup_{\substack{\Facet\in\dim' \\ \state\in\State}}
\spotMany(\Facet,\state)
& (\textrm{Associativity of\ }\cup) 
\\[1ex]
&= \iMag[\dim,\State]\cup\iMag[\dim',\State] 
& (\textrm{Definition of \iMagName})
&\qed
\end{align*}

\setlength{\abovedisplayskip}{\saveskip}





\subsection{Example}


Suppose an intersectional population of interest is low-socioeconomic women. 
This population lies at the intersection of two diversity dimensions, Gender and SES (socioeconomic status):
\begin{align*}
\dim  &= \Gender  &
\dim' &= \SES \\
\intertext{producing these definitions:}
\textrm{GenderMag} &= \iMag[\Gender] &
\textrm{SESMag} &= \iMag[\SES] &&
\end{align*}
Substituting these values into Theorem \ref{th:comp} says that the union of results of designers using the existing GenderMag~\cite{burnett2016gendermag-jrnl} and then using an emerging method, SESMag~\cite{Hu2021Toward}, produces the same results as the designers would get by using a (hypothetical) new intersectional method ``GenderSESMag''. 
\[
\iMag[\join{\Gender}{\SES}] =
\iMag[\Gender]\cup \iMag[\SES]
\]
or, in other words,
\[
\textrm{GenderSESMag} = \textrm{GenderMag}\cup \textrm{SESMag}
\]
Note that this example, motivated by the intersectional population of low-SES women, also affords analysis of the other three intersections in this combination of diversity dimensions---low-SES men, high-SES women, and high-SES men. 
Issues likely to especially impact low-SES women will emerge from designers' analysis using ``minimum'' facet values for each facet type; for high-SES men using ``maximum'' facet values for each facet type; and so on. 
Fixing identified inclusivity issues in ways that simultaneously serve both the minima and maxima of each facet type also serves people with a mix of these facet values or with values between them.  

Although this example employs two dimensions, extending to additional dimensions is straightforward for any dimension for which an \iMagName\ method exists.
For example, extending this example to a three-dimensional intersectional population (e.g., elderly low-SES women) follows the above by simply applying \joinName\ to a third diversity dimension, $\dim''=\Age$, consisting of the facet types that researchers have investigated for AgeMag~\cite{mcintosh2021agemag}. 
The number of dimensions is unlimited, so additional dimensions could continue to be added (e.g., race, country of residence, and so on) to allow increasingly focused intersectional sub-populations.

%% file: content/5_discussion.tex
\section{Discussion and Limitations}

\subsection{How can this work?}
At first glance, the proof we present may seem counterintuitive.
Years of intersectionality research seem to show exactly the opposite---that considering different diversity dimensions in isolation (e.g., race alone, or gender alone)---misses phenomena occurring in the intersections.
However, those findings have been in the context of \textit{empirical} work, which rests on only the finite samples of data that have been gathered, which has traditionally been in unequal amounts for different intersectional populations.
The key to our approach's validity is that it rests upon reasoning about \textit{all possible values}
of the facet types, which gives equal representation to populations in the intersections.


\subsection{What type-based reasoning can provide, at what price}
In the realm of programs, abstracting from individual values to reason with types allows whole categories of problems to be found (e.g., multiplying a string with an integer), without having to check particular string or integer values.
The analogous technique in analytical HCI is to reason across a range of traits of an identity, such as reasoning about the influence of skin color (facet type) rather than about individual skin colors (facet values).
Doing so can enable detection of problems on first principles; like type-based analysis, it can  greatly reduce the need to individually check every possible value of that trait.  
\NOTE{maybe we should use a facet type in imag for this instead of skin color}

\boldification{There's a price for this}
\NOTE{should it be reasoning at the type level...I dont understand how you can reason at type-abstraction...we can use type abstraction and then reason at the types?}

However, reasoning at the type-abstraction level does come at a price: We generally lose precision, since so many different values are represented by the same type. 
Fortunately, this doesn't pose a problem in our case, because the employed type abstraction preserves the relevant property (of causing an error) for all values:
Whenever analyzing a collection of individual values would report an error (that is, inclusivity bug) for some specific value, the type-based analysis will report the same error for at least one of the facet type's two extreme values. Thus, for ordinal facet types satisfying the assumptions of Section~\ref{sec:formal}, the type-based analyses can't miss an error. 
Erring on the safe side, the type-based analysis might report an error when the value-based analysis might not, but this is exactly the behavior we need because when an error is found for one of the extreme values, it is a relevant error that could be missed by a value-based analysis that did not happen to test with that value.

As with other analytical methods, our approach is a complement to empirical methods.
Empirical methods can capture issues that analytical methods miss, and analytical methods can capture issues that empirical methods miss.
Thus, design teams would be well-served to choose strategically when to use an analytical method, when to use an empirical one, and when to use both.


\subsection{Completeness} 

\boldification{The approach assumes intersection has no new facet TYPES; what about cases where there are?}
Our approach for combining methodologies depends on the completeness of the set of facet types in the underlying InclusiveMag-generated methods. 
For example, neither GenderMag nor SESMag claim a complete set of facet types---instead, to support designer efficiency, both used a subset of facet types that could have been included~\cite{burnett2016gendermag-jrnl, Hu2021Toward}.  
An intersectional composition of these methods would operate on only these facet types.
Thus, although the composition is complete for all \textit{values} of the facet types involved, it may not be complete in the set of facet \textit{types}.


%

\boldification{If a missing facet type is found: add it to a persona; consider exporting it back the original iMag}

If designers or researchers  become aware of a new facet type needed for an intersectional population, through their or others' empirical work, they can simply perform an analysis for this additional facet type to identify any new issues. 
Theorem \ref{th:comp} guarantees that the result is the same as re-performing a previous analysis with the added facet type. They should also weigh adding the facet type to one or both of the non-intersectional methods in their future work, if they believe it was a previously unknown facet type that indeed is associated with that identity.

\boldification{Shroff2011's low-SES women were a distinct intersection, but their facet types mostly match gendermag and ses-mag}

As a thought exercise, we mapped Shroff and Kam's results from an investigation of low-SES women in India~\cite{Shroff2011} to the GenderMag and SESMag facet types.
Shroff and Kam~\cite{Shroff2011} described a five-stage scale of growth towards independence among the women in their study; along these stages the women varied in such traits as their passivity vs. empoweredness, their access to resources (money, personal property), availability of emotional support, degree of literacy, and awareness of rights.  Their circumstances are particular to low-SES women in a particular cultural environment, and could not have been predicted from investigating women alone or low-SES people alone.  However, of the 7 facet types we could derive from Shroff and Kim's descriptions, all but one were covered by the union of the GenderMag and SESMag facet types (Table~\ref{tab:ShroffFacets}).

\input{tables/ShroffFacets}

\subsection{The broader context}
Our work can be viewed from the perspectives of three contrasting intersectionality research philosophies. 
%
The \emph{Anticategorical Complexity} philosophy rejects social categorizations entirely, advocating instead for an ethnographic, acategorical reporting style~\cite{mccall2005complexity, schlesinger2017intersectional}.
Dourish and Mainwaring exemplify this approach by proposing to abandon the term \emph{user} because it tends to pigeonhole all users as belonging to a single, homogeneous group~\cite{dourish2012ubicomp}.
Our work subscribes to the notion of abandoning homogeneous notions of ``the user,'' but does not entirely reject social categorizations.

In contrast, the \emph{Intracategorical Complexity} (``within-category'') philosophy acknowledges social categories, but focuses on within-category variation~\cite{mccall2005complexity, schlesinger2017intersectional}. 
An example is Woelfer et al.'s investigation of social network websites usage by homeless youth, focusing on  differences among each of the recruited participants without prior assumptions of these participants also being in categories other than ``homeless youth''~\cite{woelfer2012homeless}.
Our approach is consistent with the notion of intracategorical variation, but does not highlight it particularly. 

The third philosophy, and the one in which our work falls directly, is the \emph{Intercategorical Complexity}  (``between-categories'') philosophy.  
This philosophy acknowledges different social categories that may be intersected, and illustratively reports them (e.g., studying  differences and similarities in CS-related experiences of Black women, Black men and non-Black women) ~\cite{mccall2005complexity, ross2020intersection}. 
Another example of the usage of Intercategorical Complexity philosophy in HCI is Ames et al.'s work on the differences and similarities between middle-class and working-class families in their technology choices and usage~\cite{ames2011understanding}.
%
%
Our work aims to support researchers and practitioners to reason explicitly about inclusivity issues arising in such intersections between different categories.



%% file: tables/ShroffFacets.tex
\begin{table}[ht]
    \centering
    \small
    \begin{tabular}{|p{5.8 in}|}
    \hline
    \textit{1. At this stage, the women are still passive in that they listen and absorb the information that NGOs present, but do not actively ask questions or act on this information.} \\
     \textbf{SESMag}: Perceived control and attitude toward authority \\
    \textit{2. There is minimal emotional support for women; they cannot freely vent their problems to their husbands and parents-in-laws.} \\
     \textbf{(no equivalent)} \\

\textit{3. Women in passive stages [...] were also less at ease with technology, and hence we spent more time demonstrating the prototypes to them.} \\
    \textbf{GenderMag}: Computer self-efficacy \\

\textit{4. Women ... either cannot afford items such as cellphones, or risk having their cellphones snatched by a male family member if he is annoyed by her use of it.} \\
    \textbf{SESMag}: Access to Reliable Technology \\

\textit{5. The fear of technology is commonly reported in the HCI4D literature, especially among the lowly educated, for reasons such as nervousness about damaging the device.} \\
    \textbf{SESMag}: Attitude toward Risk, Communication Literacy/ Education/ Culture \\
    
\textit{6. Women are unaware of their rights... around child rearing, education, health, family planning, sanitation and other developmental topics.} \\
     \textbf{SESMag}: Communication Literacy/ Education/ Culture \\
    
\textit{7. It is easy to forget the details of what was covered, and since the women are semi-literate, they cannot take notes during workshops.}  \\
    \textbf{SESMag}: Communication Literacy/ Education/ Culture  \\
 \hline 
    \end{tabular}
    \caption{A mapping of the needs of low-SES women in India~\cite{Shroff2011} to facet types in GenderMag or SES-Mag. (Note: two of the above facet types (Computer self-efficacy and Risk) are common to both methods; here we attribute them to the method most likely to have spotted the issue. e.g., if the quote emphasizes women we attribute to GenderMag and if it emphasizes SES we attribute to SESMag.)}
    \label{tab:ShroffFacets}
\end{table}

%% file: content/6_implications.tex
\newcommand{\bwd}{\ensuremath{\leftarrow}}
\newcommand{\fwd}{\ensuremath{\to}}

\section{Analytical intersectionality in practice: Four use-cases}



How might HCI practitioners make practical use of this result? %
Practitioners can apply the theorem ``forward'' (in the   split/decomposition direction) to create a plan for analysis.
Then, the practitioners can apply the theorem ``backward'' (in the join/composition direction) to accumulate results of the above analyses.
%
These applications of the theorem give rise to (at least) four practitioner use-cases.



\textbf{Use case \#1: Intersectional evaluations}: 
An obvious application of the theorem is that HCI practitioners can evaluate their prototypes one diversity dimension at a time, while still covering all interesectional populations.
For example, a practitioner trying to address the needs of elderly, low-SES women, can analyze their prototype via AgeMag~\cite{mcintosh2021agemag}, then via SESMag~\cite{Hu2021Toward}, then via GenderMag~\cite{burnett2016gendermag-jrnl}.
The practitioner can then compose (take the union of) the issues found separately with AgeMag, SESMag, and GenderMag, to obtain inclusivity issues that elderly women in low socio-economic situations will face, as well as those faced by people in the other intersections of these three diversity dimensions.\footnote{Of course, differences in practitioners' HCI abilities can affect the completeness of the results obtained.  For example, these three methods use specialized cognitive walkthroughs, which are generally very strong at avoiding false positives, but somewhat weak at avoiding false negatives~\cite{mahatody2010}. However, human error is neither more nor less at play with this method than with any other endeavor conducted by HCI practitioners.}
Accumulating the results in this way is justified by applying the theorem backward.

%

\textbf{Use case \#2: Divide-and-Conquer by a team for complex evaluations}: 
By the same token, a team of HCI practitioners could divide-and-conquer a multi-dimensional analysis by splitting it into different dimensions, with different subteams doing the analysis on the different dimensions.
For example, one subteam might do the AgeMag analysis while another does the GenderMag analysis, and so on.
In fact, a team could even divide up the work of a single \iMagName\ analysis.
For example, one subteam might do an analysis in terms of only the first three facet types, with another subteam focusing on the remainder of the facet types. 
Doing so might help ward off the facet popularity phenomenon, in which a team habituates to considering certain facet types and tends to neglect others (as reported in~\cite{burnett2016gendermagFieldstudy}). 

Splitting a complex analysis into several simpler analyses that are run independently of one another, followed by composing the results is justified by applying the theorem forward and then backward.

%

\textbf{Use case \#3: Combining facet-based artifacts from multiple diversity dimensions}:
%
Some HCI practitioners have built formative and thought-exercise artifacts directly from \iMagName\ facet types and/or facet values. 
For example, HCI practitioners at some companies have used a GenderMag ``facet survey'' to gather formative data about the facet-value composition of their customer bases~\cite{anderson2021diversityAI, Hilderbrand_2020gendermagPractices, vorvoreanu2019gendermagEmpirical}.
Combining such survey questions from more than one diversity dimension---i.e., the union of such questions---adds to the length of the survey at a rate proportional to the number of facet types, which may be acceptable for some surveys.

Some HCI practitioners have also used facet types and values directly as the basis for  design heuristics and design examples specialized to one or a few diversity dimensions (e.g.,~\cite{GenderMag-designCatalog, letaw2021online, mendez2019inclusivemag, microsoftToolkit:Inclusive}).
HCI practitioners can combine such heuristics and design examples as a straightforward union operation of the facet types that define each of the dimensions of interest.  

Trust in the validity of all these practices is justified by the backward or compositional direction of the theorem.

\textbf{Use case \#4: Invite More Personas to Design Meetings}:
Some uses of \iMagName\ methods revolve around the personas that have been created around the facet types (recall Figure~\ref{figure:inclusiveMagProcessDiagram}). 
For example, following the recommendations of persona researchers such as Adlin and Pruitt, one persona-based practice is HCI practitioners ``inviting'' their personas to design meetings~\cite{adlin2010essential}, such as seating pictures of each persona around the meeting's conference table, and considering the persona's (expected) opinion as design decisions are being made.
This practice is easily expanded by inviting two personas from every diversity dimension of interest.
For example, instead of inviting just the two personas that together capture all the maximum and minimum values of the facet types in one dimension, invite two personas like these from every diversity dimension of interest.  

The addition of persona pairs to a design session is equivalent to performing  multiple analyses for different dimensions in parallel and combining the results from all analyses just in time. Formally, this scenario is equivalent to the divide-and-conquer scenario. In fact, it could be considered an extreme version of divide-and-conquer in which each individual facet type from any dimension is covered for each state of the software use case.

\boldification{MMB: this is intended to wrap it up.  Probably not the best, but perhaps good enough?}
\textbf{Beyond these four}: Additional use-cases relating and leveraging the above are also possible, such as optimizations to avoid analyzing a facet type shared by multiple dimensions multiple times, leveraging past analyses of some diversity dimension (e.g., Gender only and Age only) to obtain new insights into intersectional populations, and using facet surveys to recruit empirical populations that give equal empirical voice to every intersectional population. 

%% file: content/7_conclusions.tex
\section{Conclusions}

\boldification{}
Our work is one of those that harvests certain synergies between notions from programming languages and notions from HCI~\cite{bogart2012naturalProgramming,  chasins2021pl, coblenz2021pliers, erwig2002applesOranges, hempel2016semi, ko2008debuggingReinvented,  mccutchen2020elastic, myers2016programmersAreUsers}.
However, most such works focus on benefiting either  programming language designers or end users of a system  (e.g., individual programmers or spreadsheet users). 
In contrast, our work aims to benefit HCI practitioners---especially those interested in supporting intersectionality.
 
Toward that end, we have drawn from the programming language notion of type abstractions to create a (de)com\-po\-si\-tion\-al model enabling a population's diverse identity properties to be joined and split.
%
The model enables HCI researchers and practitioners to systematically reason about and design for intersectional populations.  
Specifically: 
\begin{itemize}
    \item The model enables analytical reasoning about facet types representing traits for which individuals at opposite ends of a diversity identity dimension differ. 
    
    \item The model allows the systematic decomposition and composition of diversity dimensions to facilitate flexible analysis strategies, which provably produce the same results.
    
    \item The model's theoretical properties ensures bringing ``equal voice'' to every intersection of the identities of interest. 
\end{itemize}
Ultimately, we hope this work can bring intersectional HCI to mainstream HCI practices.
In earlier days of HCI, many software companies believed they could not ``afford'' to do HCI~\cite{aydin2011cost-justify, donahue2001usability, dray1994cost}.
However, analytical inspection methods like Heuristic Evaluation and Cognitive Walkthroughs enabled even small companies without dedicated staff to begin to affordably engage in HCI practice.
We believe that the method we have presented for analytically harnessing the power of type abstraction can bring similar affordability to the practice of intersectional HCI.